\documentclass[12pt,preprint]{aastex}

\shorttitle{Direct Evidence for a Polar Spot on SV Cam}
\shortauthors{S.V. Jeffers et al.}

\begin{document}

\title{Direct Evidence for a Polar Spot on SV Cam}

\author{S.V.Jeffers\altaffilmark{}, A.Collier
Cameron\altaffilmark{} J.R.Barnes\altaffilmark{}}
\affil{School of Physics and Astronomy, University of St Andrews,
North Haugh, St Andrews, Fife KY16 9SS, U.K.}
\email{svj1@st-and.ac.uk}

\author{J.P.Aufdenberg\altaffilmark{1}}
\affil{National Optical Astronomy Observatory, 950 N. Cherry Ave,
Tucson, AZ 85726, U.S.A.}

\and

\author{G.A.J.Hussain\altaffilmark{1}}
\affil{Astrophysics Div., Research \& Science Support Department of ESA,
ESTEC, Postbus 299, Noordwijk, The Netherlands}

\altaffiltext{1}{Harvard-Smithsonian Center for Astrophysics, 60 Garden Street,
Cambridge, MA 02138 U.S.A.}

\begin{abstract}

We have used spectrophotometric data from the Hubble Space Telescope
to eclipse-map the primary component of the RS CVn binary SV Cam over
9 HST orbits.  We find from these observations and the {\sc hipparcos}
parallax that the surface flux in the eclipsed low-latitude region of
the primary is about 30\%\ lower than computed from a {\sc phoenix}
model atmosphere at the effective temperature that best fits the
spectral energy distribution of the eclipsed flux.  This can only be
accounted for if about a third of the primary's surface is covered
with unresolved dark star-spots.  Extending this to the full surface
of the primary, we find that even taking into account this spot
filling factor there is an additional flux deficit on the primary
star.  This can only be explained if there is a large polar spot on
the primary star extending from the pole to latitude 48$\pm$6$^\circ$.

\end{abstract}


\section{Introduction}

Over the last 20 years, Doppler imaging studies have revealed that
rapidly rotating RS CVn binary systems frequently show long-lived
polar caps.  These features have been shown to be quite stable over
extended time-scales of up to several years.  Examples include HR 1099
\citep{vogt99hr1099,petit02hr1099}, EI Eri \citep{hatzes92eieri,washuettl01}, 
UZ Lib \citep{olah02uzlib}, HU Vir \citep{strassmeier94,hatzes98},
V1355 Ori \citep{strassmeier00} and IL Hya \citep{weber98ilhyd}.
Theoretical efforts to understand this phenomenon are currently
focused on the role of meridional flows in sweeping magnetic flux from
decaying active regions up towards the poles
\citep{schrijver01polar}. In order to produce a polar spot in these
simulations, however, bipolar active regions have to emerge at a rate
30 times faster than solar, implying that the photospheres of active
stars should be peppered with many small spots.

In order to test this idea by measuring this filling factor, we were
awarded 9 orbits of HST/STIS time in 2001 November to eclipse-map the
starspot distribution on the F9V primary of SV Cam.  We obtained
spectrophotometric light-curves through three primary eclipses with a
photometric signal-to-noise ratio of 5000.  At this unprecedented
precision, we discovered that using conventional limb-darkening models
based on plane-parallel atmospheres gave very poor fits to the primary
eclipse profile, particularly around the contact points
\citep{jeffers04c}.  We concluded that the spiked pattern of residuals
could only be accounted for by a combination of incorrect binary system
parameters, and the assumed presence of a polar cap on the primary
star.

SV Cam has been the target for numerous spectroscopic and photometric
observations.  Despite this, there are no established values for the
principal stellar parameters that define the shape of the photometric
lightcurve i.e. the temperatures and radii of the primary and
secondary stars.  As shown in \citet{jeffers04c}, the determination of
these parameters is essential for a correct modelling of the binary
system light-curve.  If the system parameters are wrong a satisfactory
fit to the lightcurve will only be obtained by placing a greater total
spot area than is present when the correct parameters are used.  In
\citet{jeffers04c} we determined the radii of the two stars through
lightcurve fitting, though it was not possible to model the depth of
the eclipse correctly without artificial spot features appearing on
the final image.  However, when an assumed polar cap was also included
in the model, the artificial spot features also disappeared.  It is not
possible to reconstruct a surface brightness distribution for all
latitudes as the poles of SV Cam's primary remain uneclipsed.

In this paper we empirically determine the temperatures for both
primary and secondary components by fitting {\sc phoenix} spherical
model atmospheres to our spectra.  As our HST data is calibrated in
absolute flux we determine how much light is missing due to the
surface of SV Cam's primary being covered with small star-spots.
Finally we show that there are two polar caps on the primary star and
determine their size.

\section {Observational Details}

\begin{figure}
\includegraphics[angle=270,scale=.40]{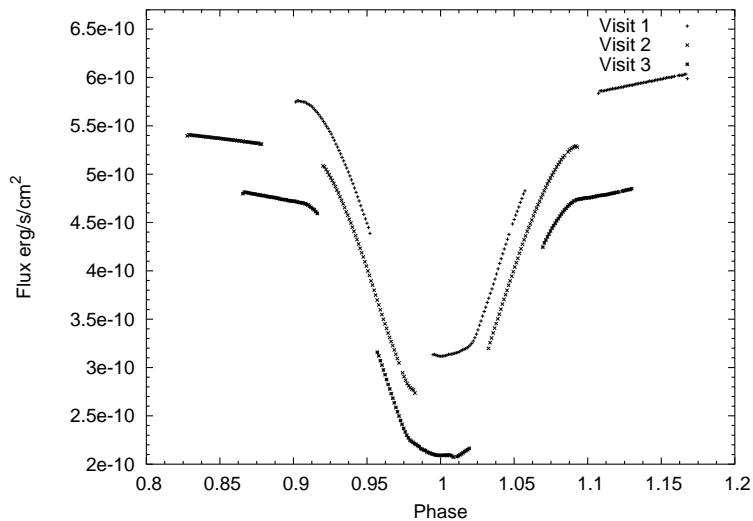}
\caption{The 9 spacecraft orbits that comprise the 3 primary eclipses
of SV Cam (where Visit 1 is at the top).  An offset of 5$\times$$10^{-11}$ has been included for clarity.}
\label{f-obs}
\end{figure}

Three primary eclipses of SV Cam were observed by the HST, using the
Space Telescope Imaging Spectrograph with the G430L grating.  The
observations comprised 9 spacecraft orbits and spanned 5 days at 2 day
intervals from 1-5 November 2001 as shown in Table~\ref{hstobs28}.
Summing the recorded counts over the observed wavelength range
2900\,\AA\, to 5700\,\AA\ for each spectra, results in a photometric
lightcurve.  A plot showing the 9 spacecraft orbits and the 3 primary
eclipses that comprise the photometric lightcurve are shown in
Figure~\ref{f-obs}.  The observations have a cadence of 40s and a
photometric precision of 0.0002 magnitudes (S:N 5000) per 30\,s
exposure. These observations and the data reduction method are
explained in greater detail in
\citet{jeffers04c}.

\section{Phoenix Model Atmospheres}

The synthetic spectra we use in this paper is based on an extension of
the grid of {\sc phoenix} model atmospheres described by
\citet{allard00}.  This extended grid includes surface gravities,
$\log(g) > 3.0$ needed for main sequences stars.  These models are as
described by \citet{hauschildt99}, but include an updated molecular
line list.  The models are computed in spherical geometry with full
atomic and molecular line blanketing using solar elemental abundances.
In these models, the stellar mass is 0.5 M$_\odot$ and the convection
treatment assumes a mixing-length to pressure scale height ratio of 2.
There are 117 synthetic spectra in total.  The effective temperature
runs from 2700K to 6500 K in 100K steps at three surface gravities:
$\log(g)$ = 4.0, 4.5, and 5.0.  The wavelength resolution of these
synthetic spectra is 1\AA.

The {\sc phoenix} spectra were converted to the same resolution scale
as the HST spectra by convolving them with a Gaussian profile of 
width equal to the instrumental resolution, 2.73\AA.  Each spectrum was then
trimmed to the wavelength range 3000\AA\ - 5700\AA.

\section{Temperature Fitting using Hipparcos Distance}

In this section we determine the temperature for both the primary and
secondary components of SV Cam using {\sc phoenix} model atmospheres.
For consistency we shall denote the flux at the Earth by $f$, and the
flux at the star by $F$.

\subsection{Primary Star + Secondary Star}

In \citet{jeffers04c} we accurately determined the radii of the
primary and secondary components through lightcurve fitting to be
1.23\,$\pm$0.03\,R$_{\sun}$ and 0.79\,$\pm$0.02\,R$_{\sun}$ for the
primary and secondary stars respectively.  Knowing these radii we can
evaluate the flux contribution from the secondary star relative to
that of the primary star. The {\sc phoenix} spectra are scaled by
$r^2/d^2_{HIPP}$ to give flux at the Earth rather than at the stellar
surface.  The {\sc phoenix} flux received at the Earth outside of
eclipse can be expressed as;
 
\begin{equation}
f_{ptotal}=\frac{r^2_{pri}}{d^2_{HIPP}}(F_{pri} + \frac{r^2_{sec}}{r^2_{pri}}F_{sec})
\label{eq-ftot}
\end{equation}

where $F_{pri}$ and $F_{sec}$ are the {\sc phoenix} model atmosphere
fluxes for the primary and secondary stars respectively; $r_{pri}$ and
$r_{sec}$ are the primary and secondary radii, $d_{HIPP}$ is the {\sc
hipparcos} distance (84.96\,$\pm$\,8.5\,pc).  We fitted this equation
for each permutation of primary temperature in the range $T_{pri}$ =
4500\,K to 6500\,K, and secondary temperature $T_{sec}$ = 2800\,K to
5500.  The best fitting pair of temperatures was determined using
$\chi^2$ minimisation;

\begin{equation}
\chi^2 = \sum_{\lambda (i)=1}^{N} \left(\frac{f_{out}(i)- \gamma_{total} * f_{ptotal}(i)}{\sigma(i)}\right)^2
\end{equation}

where $f_{out}$ is an averaged spectrum outside of primary eclipse, and
$\gamma_{total}$ is a scaling factor.  A scaling factor is included so
that the shape of the spectrum is fitted rather than the absolute flux
levels.  This scaling factor is defined as;

\begin{equation}
\gamma_{total} = \frac{\sum f_{out} f_{ptotal}/\sigma^2(i)}{\sum (f_{ptotal})^2/\sigma^2(i)}
\label{e-scale1}
\end{equation}

\begin{figure}
\hspace{0.1cm}
\includegraphics[angle=270,scale=.40]{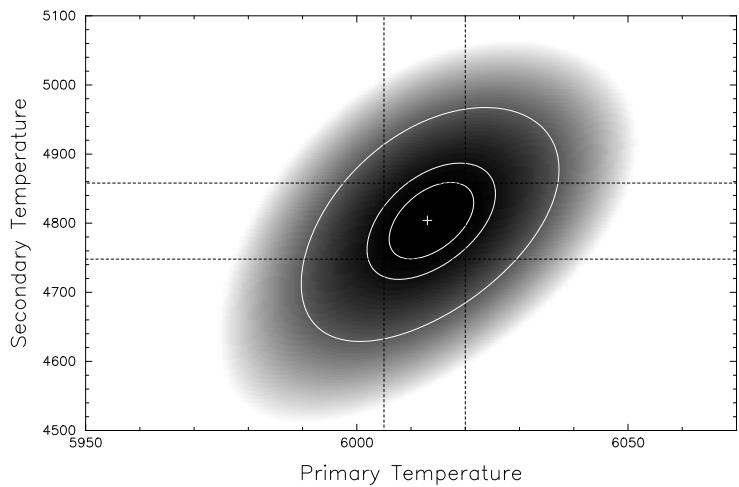}
\caption{Contour plot of the $\chi^2$ landscape of the combined primary and 
secondary stars.  The minimum $\chi^2$ value occurs at
6013\,$\pm$\,19\,K, and 4804\,$\pm$\,143\,K, for the primary and
secondary stars respectively.  From the centre of the plot the first
contour ellipse represents the 1 parameter 1\,$\sigma$ confidence
limit at 63.8\%, the second ellipse represents the 2 parameter
1\,$\sigma$ confidence limit at 63.8\% whilst the third elipse
represents the 2 parameter 2.6\,$\sigma$ 99\% confidence limit.}
\label{contour}
\end{figure}

\begin{figure}
\includegraphics[angle=270,scale=.40]{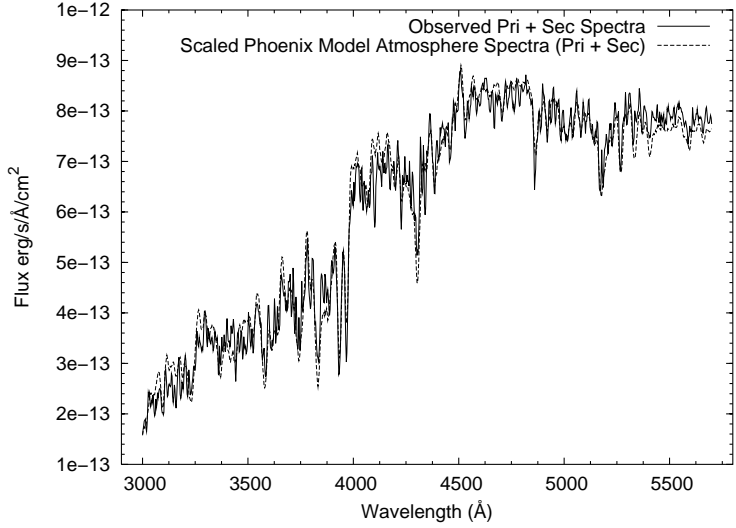}
\caption{Plot showing the combined primary and secondary stars outside of eclipse, $f_{out}$, and the best fitting {\sc phoenix} spectrum (6000\,K and 4800\,K for the primary and secondary stars respectively).}
\label{phxps}
\end{figure}

The results for the $\chi^2$ minimisation are shown in
Figure~\ref{contour} in the form of a contour plot.  The minimum
$\chi^2$ value occurs at 6013\,$\pm$\,19\,K, and 4804\,$\pm$\,143\,K,
for the primary and secondary stars respectively.  As is evident from
the shape of the contour plot and the size of the errors for each
temperature, the primary temperature is very sensitive, in contrast to
the secondary temperature, to small temperature changes (i.e. the
secondary places a lower constraint on the solution).  The best
fitting {\sc phoenix} model atmospheres, 6000\,K and 4800\,K, and an
observed spectrum outside of eclipse are shown in Figure~\ref{phxps}.
We neglect any reddening correction due to the proximity of SV Cam.

\subsection{Primary Star}

The first step in determining the temperature of the primary star in
isolation is to remove its spectrum from that of the secondary star.
To achieve this we simply subtracted an average spectrum inside the
primary eclipse, $f_{ecl}$, from an average spectrum outside of
eclipse, $f_{out}$.  The spectrum that results is the spectrum of the
primary star, but with a reduced flux level that is equivalent to the
primary star having a radius of the secondary star.  This missing
light will be referred to as $f_{mis}$ in the rest of this paper.
 
The contribution to the total {\sc phoenix} flux is from only the 
primary star, so  equation (~\ref{eq-ftot}) becomes;

\begin{equation}
f_{ptotal(pri)}=\frac{r^2_{pri}}{d^2_{HIPP}}(F_{pri})
\label{eq-fpri}
\end{equation}

As before, the optimal value of the primary temperature is determined
through $\chi^2$ minimisation.  The scaling factor is then equal to;

\begin{equation}
\gamma_{pri} = \frac{\sum f_{mis} f_{ptotal(pri)}/\sigma(i)^2}{\sum (f_{ptotal(pri)})^2/\sigma(i)^2}
\label{e-scale2}
\end{equation}
 
\begin{figure*}
\includegraphics[angle=270,scale=.40]{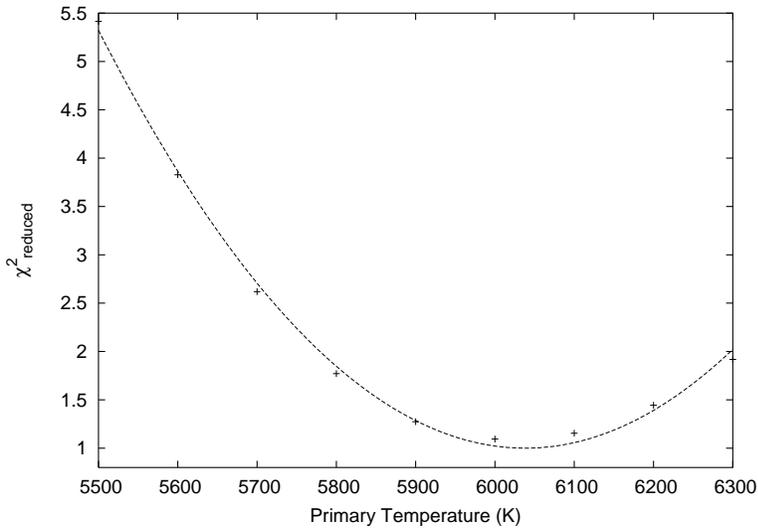}
\caption{The parabolic fit that determines the temperature minimum to be 6038 $\pm$ 58\,K}
\label{f-tempmin}
\end{figure*}

Figure~\ref{f-tempmin} shows how the reduced $\chi^2$ varies with
primary temperature.  A parabolic fit results in a minimum temperature
of 6038\,$\pm$\,58\,K.  The errors were calculated by setting
$\Delta\chi^2$ = 1 on the full $\chi^2$ value.

\section{Star-spot coverage}

\begin{figure}
\includegraphics[angle=270,scale=.40]{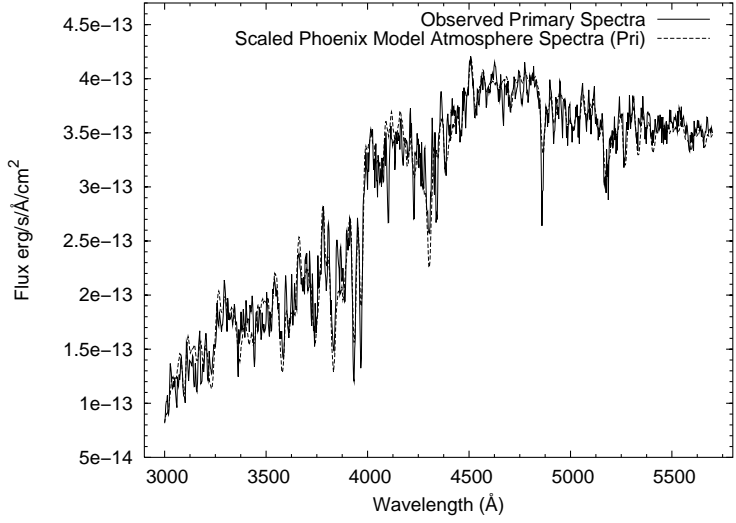}
\caption{Plot showing the spectrum
of the primary star, $f_{mis}$, and the closest match {\sc phoenix}
model atmosphere (6000\,K)}
\label{f-temp}
\end{figure}

\begin{figure}
\includegraphics[angle=270,scale=.40]{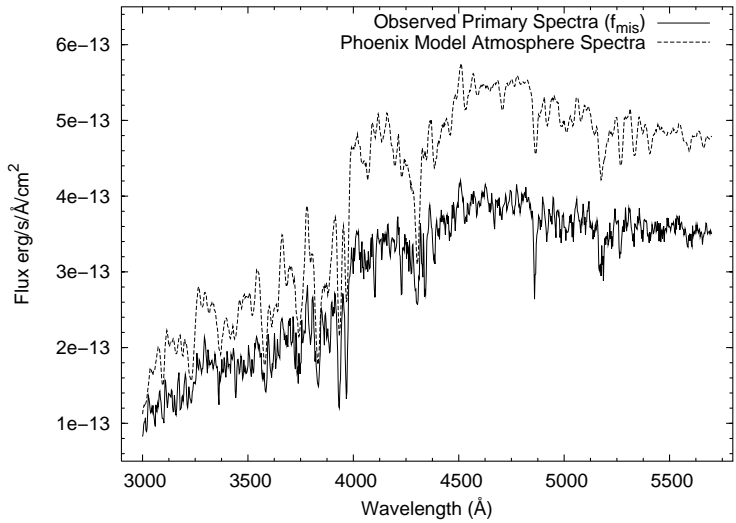}
\caption{As per Figure~\ref{f-temp} but without using a scaling
factor.This plot illustrates the flux deficit due to the presence of
dark star-spots on the surface of the primary.}
\label{prins}
\end{figure}

In Figure~\ref{f-temp} we show the best fitting {\sc phoenix} spectrum to
the primary spectrum, which include a scaling factor. The advantage
of using HST spectro-photometric observations is that for each point
on our photometric lightcurve we have a value for its absolute flux
(i.e. the flux integrated over 3000\AA\ - 5700\AA\ in each spectrum).
Replotting Figure~\ref{f-temp} without the scaling factor in
Figure~\ref{prins}, clearly shows that there is a flux deficit on the
surface of the primary star, when compared with the flux computed from
a {\sc phoenix} model atmosphere at the effective temperature that
best fits the spectral energy distribution of the eclipsed flux.

Using the eclipse mapping technique \citet{jeffers04c}, have already
verified the existence of several dark spots at low latitudes on the
primary star, but these are insufficient to account for the flux
deficit.  The total missing flux can only be accounted for if the
primary's surface is peppered with unresolved dark star-spots.  The
dark star-spot filling factor is given by:

\begin{equation}
\alpha = 1 - \gamma
\label{e-alpha}
\end{equation}
 
where $\alpha$ is the fractional star-spot coverage, and $\gamma$ is
the scaling factor as defined in equation~\ref{e-scale2}.  This
results in the fractional coverage of dark star-spots on SV Cam to be
28\%.  Taking into account the error on the {\sc hipparcos} distance
this translates to if d$_{HIP}$=93.46\,pc then the spot coverage will
be 14\%, whilst if d$_{HIP}$=76.46\,pc then the spot coverage will be
41\%.

\section{Polar Spot}

Our determination of the spot coverage factor only accounts for the
equatorial latitudes of the primary that are eclipsed by the
secondary.  However, due to the near 90$^\circ$ inclination of the binary
system it is not possible to recover high-latitude spots or polar caps
through the eclipse mapping method.  In this section we will firstly
show that there is a polar cap on SV Cam and then we will determine
its size.

\subsection{Extent of a Polar Cap on SV Cam}

If there is a polar cap on SV Cam, then total flux on the two stars
can be expressed as (following from equation(~\ref{eq-ftot}));

\begin{equation}
f_{ptotal}=\frac{r^2_{pri}}{d^2_{HIPP}}(F_{pri}(1-A_{pc}) +
\frac{r^2_{sec}}{r^2_{pri}}F_{sec})
\label{eq-ftotpc}
\end{equation}

where A$_{pc}$ is the area of the polar cap.  To obtain a full view of
the primary star's surface we used an averaged spectrum out of
eclipse, but with the secondary's contribution subtracted.  It is
valid to subtract the {\sc phoenix} spectrum as we assume that the
secondary is not spotted or faint enough that the additional flux
deficit from star-spots on the secondary is negligible.  Following from
equation(~\ref{e-scale1}) the scaling factor $\gamma$ now becomes:

\begin{equation}
\gamma_{pc} = \frac{f_{out}-\frac{r^2_{sec}}{d^2_{HIPP}}F_{sec}}{\frac{r^2_{pri}}{d^2_{HIPP}}(1-A_{pc})F_{pri}}
\label{eq-gammapc}
\end{equation}
  
If there are no polar caps on SV Cam, A$_{pc}$ will be zero and the
flux deficit should equal the previously determined flux deficit in
the primary spectrum, i.e.$\gamma_{pc}$ = $\gamma_{pri}$.  We assume
that the previously calculated spot coverage area for the missing
light of the primary is valid for the whole primary star, where
equation(~\ref{e-scale2}) is extended to use r$_{pri}$ rather than
r$_{sec}$.  Equivalently we can set $\gamma_{pc}$ equal to
$\gamma_{pri}$ to solve for A$_{pc}$, e.g.;

\begin{equation}
(1 - A_{pc}) = \frac{f_{out}-\frac{r^2_{sec}}{d^2_{HIPP}}F_{sec}}{f_{mis}}
\label{eq-areapc}
\end{equation}

\begin{figure}
\includegraphics[angle=270,scale=.40]{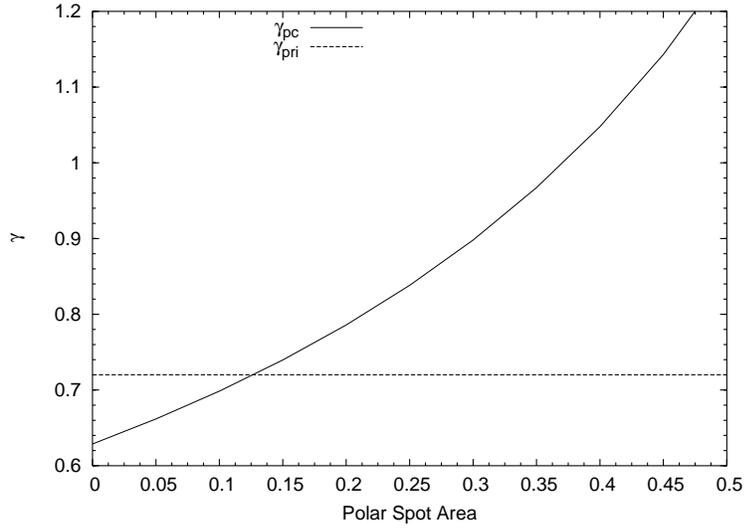}
\caption{Graphical representation of equation~\ref{eq-areapc}, where $\gamma_{pc}$ (equation~\ref{eq-gammapc}) is plotted as a function of polar spot area 
along with $\gamma_{pri}$.  The area of the polar cap is where the two
lines intersect at Spot Area=0.125.}
\label{gam-pphx}
\end{figure}

This equation is illustrated by Figure~\ref{gam-pphx}, where the
$\gamma_{pc}$ and $\gamma_{pri}$ are plotted as a function of polar
spot area.  The intersection point (0.142) indicates the fractional
area of the projected disc that is covered by polar caps on SV Cam.

\subsection{Reconstruction of a theoretical polar cap on SV Cam}

\begin{figure}
\includegraphics[angle=270,scale=.40]{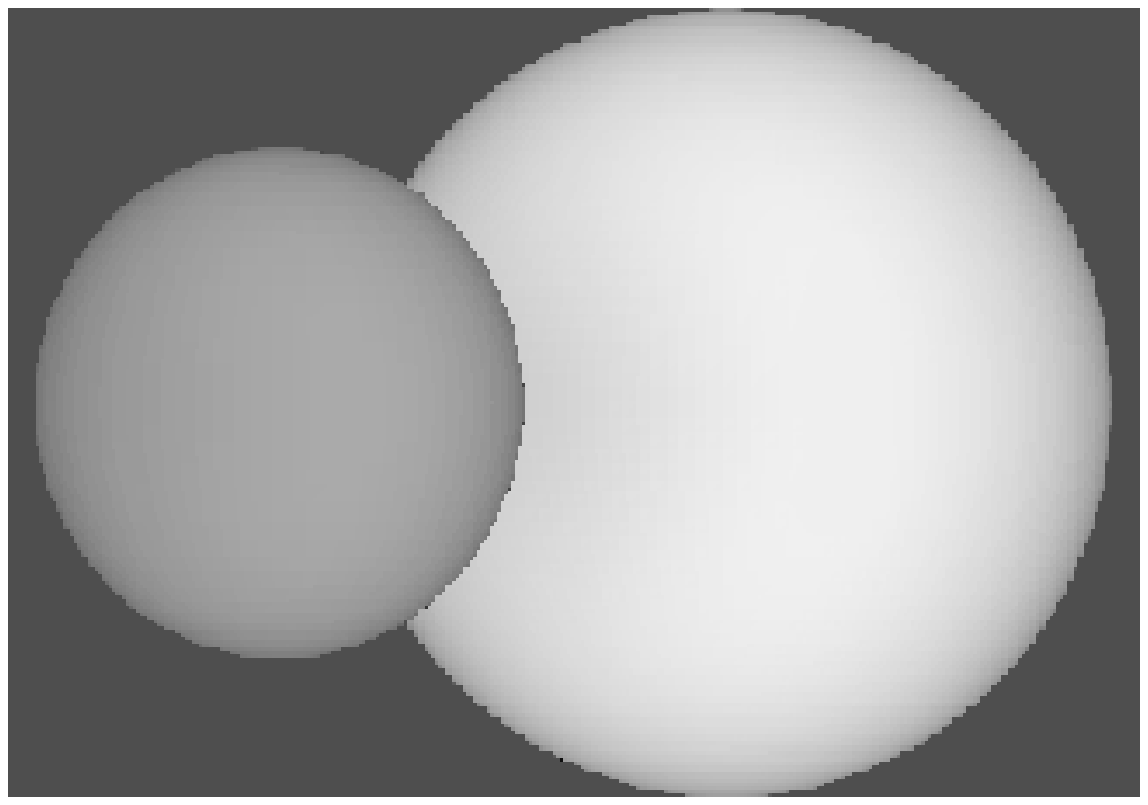}
\includegraphics[angle=270,scale=.40]{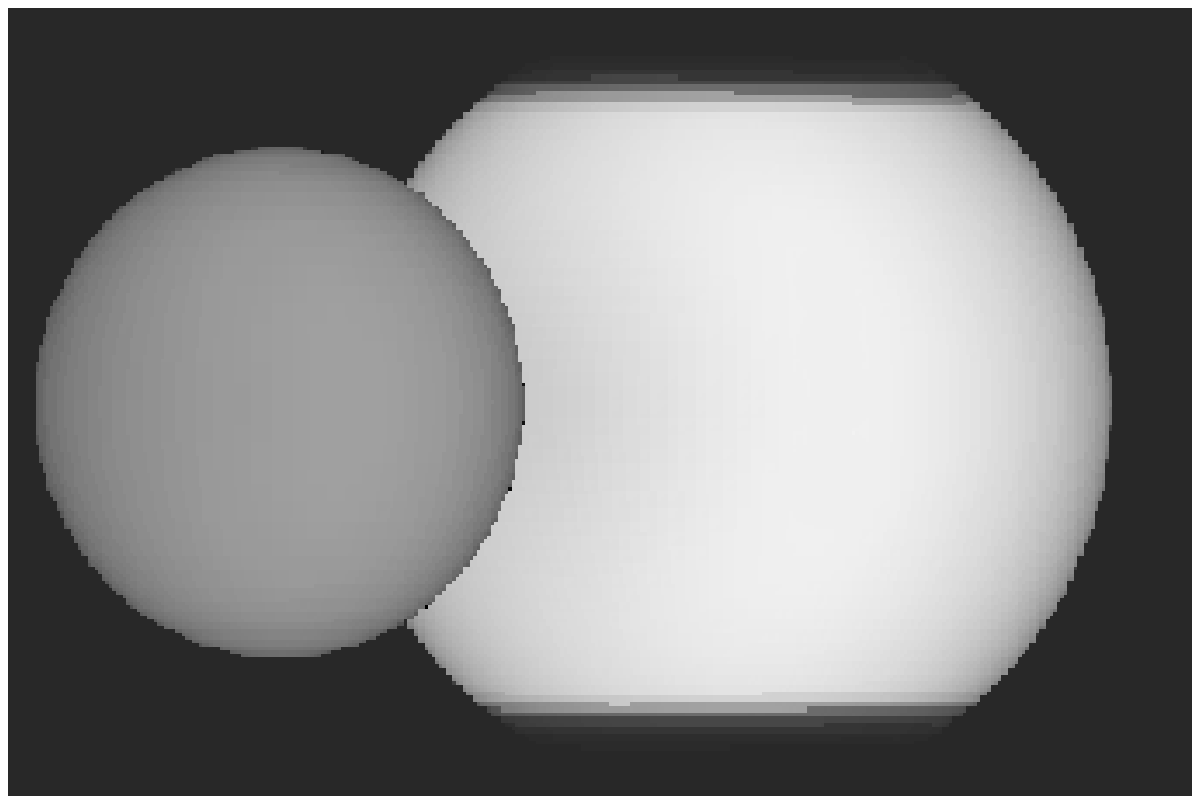}
\includegraphics[angle=270,scale=.375]{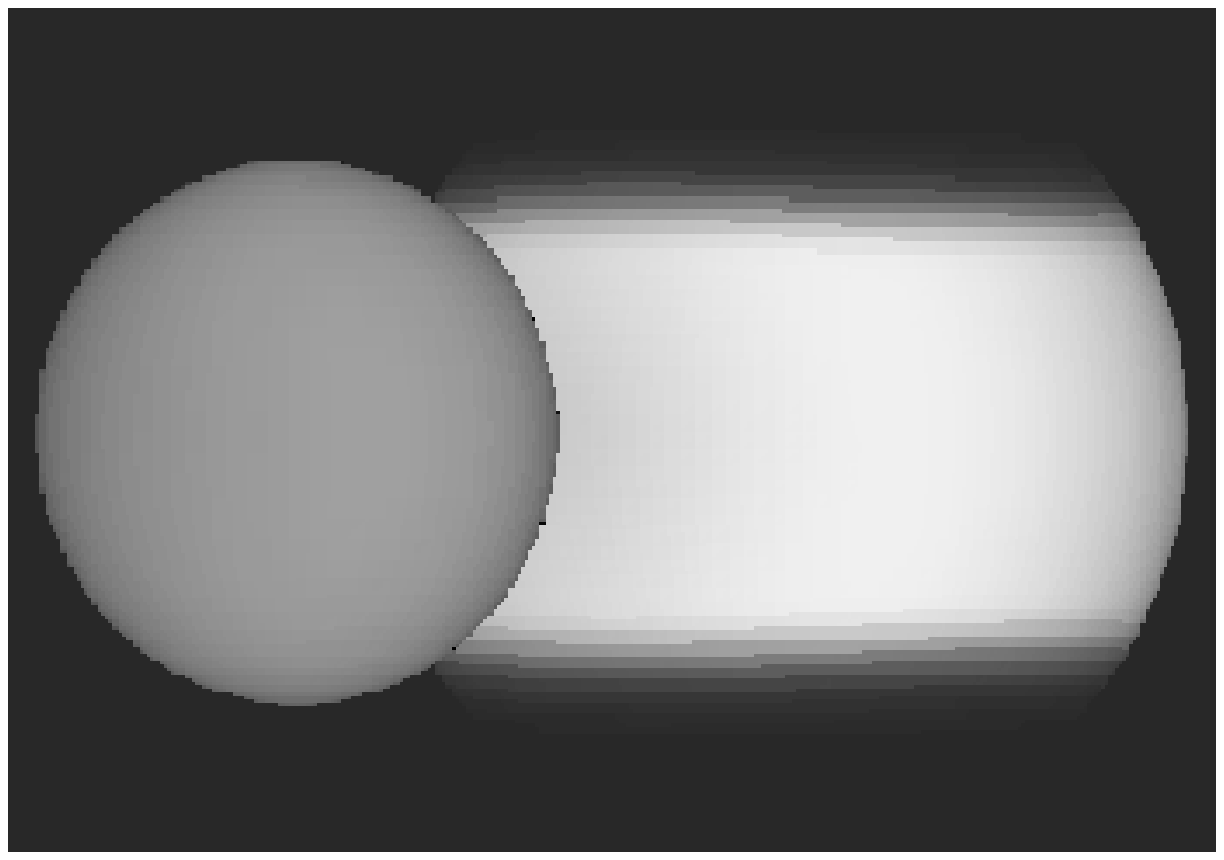}
\caption{Pictorial representation of the polar spots of SV Cam, 
with no polar cap, a 40$^\circ$ polar cap and a 60$^\circ$ polar cap.}
\label{dots}
\end{figure}

The value for A$_{pc}$ is only a fractional value by which the polar
cap has decreased the light of the star.  It does not take into
account that the star is a sphere, rather than a disc, any limb or
gravity darkening, or any spherical oblateness.  To account for these
parameters, the binary eclipse-mapping tomography code DoTs \citep{cameron97dots} was used to model theoretical polar spots on the surface
of SV Cam.  The input to DoTs is in the form of (i) longitude, set to
0$^\circ$; (ii) latitude, set to 90$^\circ$; (iii) radius set to 10$^\circ$,
20$^\circ$, 30$^\circ$,40$^\circ$,50$^\circ$ and 60$^\circ$; (iv) brightness and (v)
sharpness, which were both set to make the spot as uniformly dark as
possible.  Examples of a star with no polar spot, with a 40$^\circ$ and 60$^\circ$ polar
spot are shown in Figure~\ref{dots}.  We assume that the polar cap 
does not contribute to the total flux of the star.

\subsection{Effect of a polar cap on the primary eclipse}
 
The effect of the polar cap on the primary eclipse will be to increase
the depth of the eclipse.  The presence of a polar cap effectively
reduces the area of the primary star.  When the primary star is then
eclipsed the secondary obscures a larger fraction of the primary's
surface area than when there was no polar spot.  This is illustrated
in Figure~\ref{f-theorpc}, where there is an obvious difference in the
eclipse depth for a star with no polar cap and one with a 60$^\circ$ polar
cap.

\subsection{Size of the polar cap on SV Cam}

\begin{figure}
\includegraphics[angle=270,scale=.375]{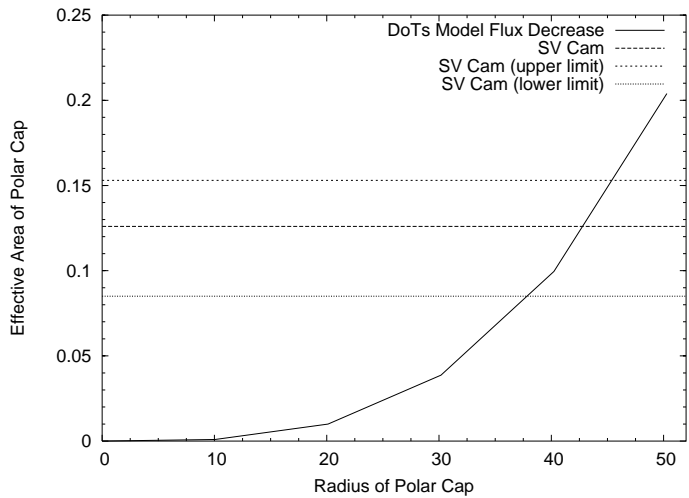}
\caption{Plot of the fractional decrease in the stellar flux of SV Cam
as a function of theoretical polar spot size radius in degrees.  Also plotted is the fractional missing light
for SV Cam as determined in the previous section.}
\label{f-pc}
\end{figure}

The fractional decrease in stellar flux as a function of theoretical
polar spot size is shown in Figure~\ref{f-pc}, where we assume that
the polar cap contributes negligible flux.  Also plotted in
Figure~\ref{f-pc} is the fractional area of a polar-spot as deduced in
the previous section.  We conclude that the polar spot on SV Cam is
42\,$\pm$\,6\,$^\circ$.

\subsection{Lightcurve of SV Cam}

\begin{figure}
\includegraphics[angle=270,scale=.375]{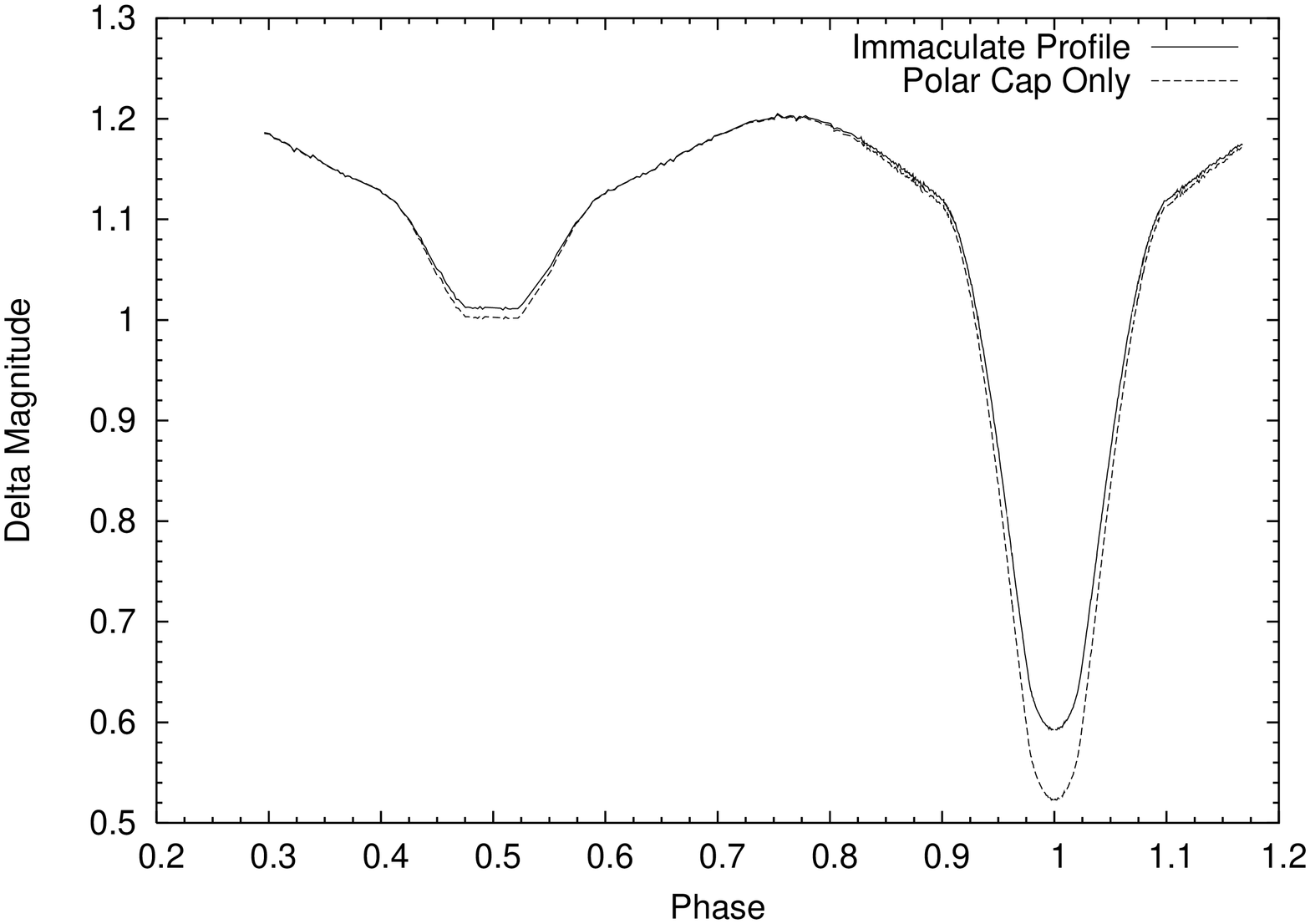}
\caption{The effect of a 42$^\circ$ polar cap on the lightcurve of SV Cam.  For comparison 
an immaculate profile, with no star-spots, is also plotted.}
\label{f-svcampc}
\end{figure}

\begin{figure}
\includegraphics[angle=270,scale=.375]{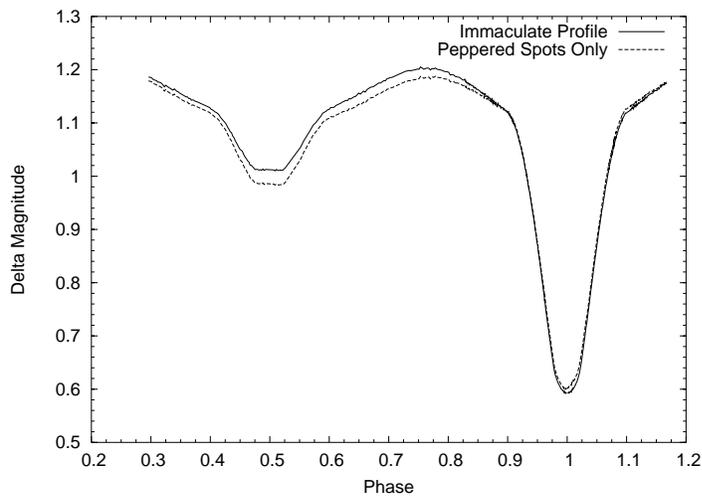}
\caption{The effect of 28\% small star-spot coverage on SV Cam. For comparison 
an immaculate profile, with no star-spots, is also plotted.}
\label{f-svcampp}
\end{figure}

As shown in Figure~\ref{f-theorpc} the presence of a polar spot will
increase the depth of the primary eclipse.  Figure~\ref{f-svcampc}
shows how a 42$^\circ$ polar cap will influence the lightcurve of SV Cam.
Contrasting this, the peppering of small star-spots will decrease the
depth of the eclipse as shown in Figure~\ref{f-svcampp}.  In a
complimentary paper, \citet{jeffers04c}, we show how to obtain the 
best fits to the data by including these effects.

\begin{figure}
\includegraphics[angle=270,scale=.375]{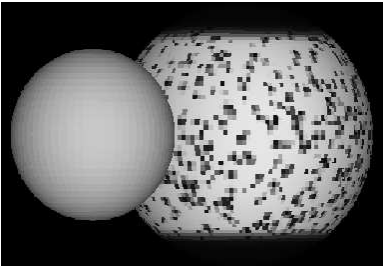}
\caption{A model of SV Cam showing 28\,\%
spot coverage and a 42$^\circ$ polar cap.}
\label{f-svcam}
\end{figure}

\begin{figure}
\includegraphics[angle=270,scale=.375]{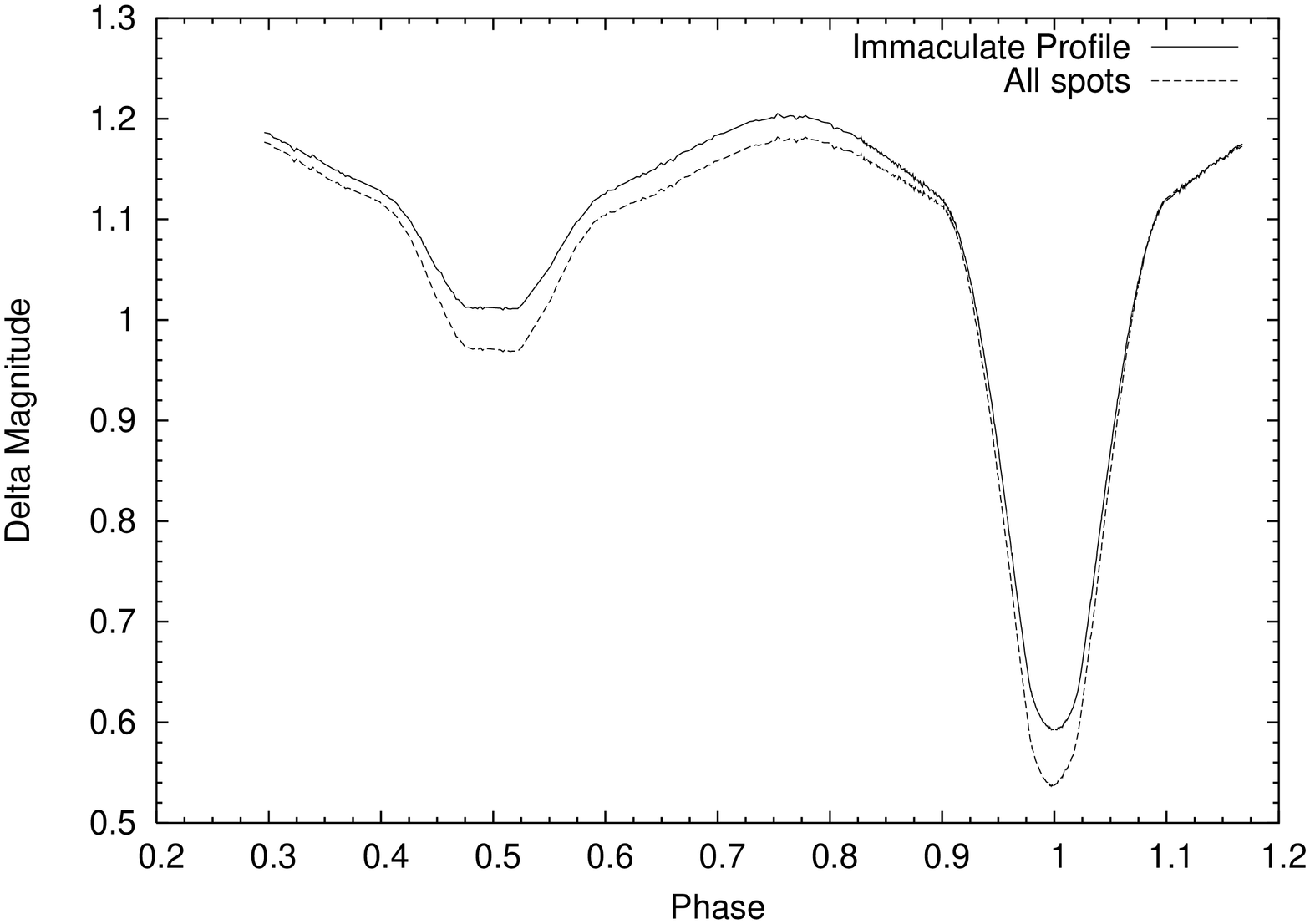}
\caption{The effect of a combination of 42$^\circ$ polar cap and 28\% spot 
coverage on the lightcurve of SV Cam.  For comparison an immaculate
profile, with no star-spots, is also plotted.}
\label{f-svcamlc}
\end{figure}

The combination of 42$^\circ$ polar caps and 28\% surface coverage of
small star-spots is shown in Figure~\ref{f-svcam}.  This translates to
the lightcurve shown in Figure~\ref{f-svcamlc}.  For clarity the
lightcurve of SV Cam is also shown without any spot features.

\section{Discussion}

The best fitting temperature of the primary star, 6013\,K, is in good
agreement with the value of 6000\,K of \citep{lehmann02}, whilst the
secondary temperature, 4804\,K, is in best agreement with the value of
4700\,K of \citep{rainger91xyuma}.  The temperatures that other
investigators have observed range from 5700\,K \citep{patkos94} to
6440\,K \citep{albayrak01} for the primary star and 4140\,K
\citep{hilditch79} to 5600\,K \citep{popper96} for the secondary star.
The range in values extends from poor data quality, incomplete
lightcurves and the inability to solve simultaneously for the binary
system parameters and the spot coverage.  However, the advantage that
we have is that we have spectrophotometric HST observations i.e. a
photometric lightcurve with absolute fluxes which can be directly
compared with {\sc phoenix} model atmospheres to fit for the
temperatures of the two stars.

There are two important results from this paper concerning the spot
coverage of SV Cam.  The first is that the surface of SV Cam is
peppered by 28\% star-spots.  \citet{jeffers04c} reconstructed an
image of SV Cam using the HST and ground based photometric data using
the eclipse mapping technique.  Our final image showed the presence of
large dark star-pots at equatorial latitudes but they were not large
enough to account for 28\% of the eclipsed part of the primary's
surface being covered with star-spots.

The second important result of this paper is that there is certainly a
dark polar cap on SV Cam.  The existence of polar caps on magnetically
active stars has always been a controversial issue.  The spectroscopic
signature of a polar cap is a line profile with a flat-bottomed core.
In some stars with low {\em vsini} this could conceivably be caused by
chromospheric in-filling of the cores of strong photospheric lines 
\citep{byrne92,byrne96}.  This result independently confirms that 
the polar cap is a reality and is not just an artifact of the imaging
process.  The high resolution spectroscopic observations of
\citet{lehmann02} showed that SV Cam had no polar cap, but only high
latitude spots.  The surface area of the primary star that has a polar
cap in this work is far greater than in these observations.  For the
extent of spot coverage that we observe, other Doppler images of RS
CVn stars would indicate that this structure is in the form of a polar
cap.
 
\begin{figure}
\includegraphics[angle=270,scale=.375]{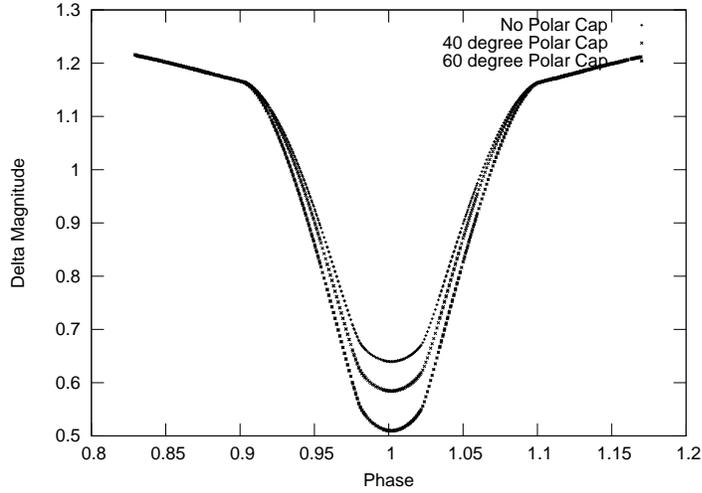}
\caption{Variation of a theoretical primary eclipse lightcurve due 
to the presence of a polar spot of radius of 40$^\circ$ and 60$^\circ$.  For
comparison the primary eclipse lightcurve without a polar spot is also
plotted.  We show only the primary eclipse as this is the part of
the lightcurve that is most effected by the presence of a polar cap.}
\label{f-theorpc}
\end{figure}

In \citet{jeffers04c} we showed that lightcurve models without a
polar cap were not sufficient to correctly fit the primary eclipse
profile particularly at the contact points.  As shown in
figure~\ref{f-theorpc}, the presence of a polar cap will increase
the depth of the primary eclipse, and if it is not taken into account
when fitting the lightcurve, it will not be possible to solve the
binary system parameters correctly.  \citet{jeffers04c} further solved
for the radii and polar cap size using $\chi^2$ minimisation.  The
optimally fitting polar cap size of 46$\pm$8$^\circ$ is in good
agreement with the value independently determined in this paper.

TiO-band monitoring \citep{oneal98tio} has indicated that between 30\%
and 50\% of a star's surface may be covered in starpots at all times.
Our result of 28\% spot coverage for the eclipsed equatorial regions
in addition to a 14\% polar cap are in agreement with TiO-band
findings.  Conventional Doppler images show that up to 20\% of the
star's surface is spotted, mainly in the form of high latitude
structure and a polar cap.  This discrepancy can be accounted for if
the star's surface is peppered with dark star-spots, too small to be
resolved with Doppler imaging.

The high total spot coverage on the primary star can have important
implications for the overall structure of the
star. \citet{spruit86spots} investigated this and concluded that over
thermal timescales the star will readjust its structure to compensate
in radius and temperature.  Also standard colour-surface brightness
relation will break down if star-spots cover a large fraction of the
stellar surface, but contribute little to the stellar optical flux.
To illustrate this point, it is possible to determine the stellar
radius, {\em R}, using standard colour-surface brightness relations to
obtain the angular diameter and the distance from the {\sc hipparcos}
parallax.  Using the relation $Rsini$=$(P_{rot}/2\pi)vsini$
\citet{odell94distance}, in a survey of axial inclination of G dwarfs
on the $\alpha$\,Per and Pleiades clusters found that $sini >$1 for
several stars.  In the context of our results this can be understood
if the high spot coverage causes $R$ to be underestimated, and would
account for the large variance of binary system parameters for SV Cam.

This work provides strong evidence, independently of Doppler Imaging,
that the poles of SV Cam's primary star are darkened by extensive
polar caps, and that at lower latitudes the photosphere is peppered
with small star-spots as has been suggested by TiO band monitoring.
It is important to establish that these stars are in fact peppered by
small star-spots in addition to the presence of a polar cap, as this
can significantly impact theoretical interpretations of the spot
distributions on these stars.

\acknowledgments

SVJ acknowledges support from a PPARC research studentship and a
scholarship from the University of St Andrews. 

JPA was funded in part by a Harvard-Smithsonian CfA Postdoctoral
Fellowship and in part under contract with the Jet Propulsion
Laboratory (JPL) funded by NASA through the Michelson Fellowship
Program. JPL is managed for NASA by the California Institute of
Technology.

Facilities: \facility{HST(STIS)}.

\bibliographystyle{mn2e}
\bibliography{iau_journals,master,ownrefs}

\begin{thebibliography}{}

\bibitem[\protect\citeauthoryear{{Albayrak}, {Demircan}, {Djura{\v s}evi{\'
  c}}, {Erkapi{\' c}} \& {Ak}}{{Albayrak} et~al.}{2001}]{albayrak01}
{Albayrak} B.,  {Demircan} O.,  {Djura{\v s}evi{\' c}} G.,  {Erkapi{\' c}} S.,
    {Ak} H.,  2001, A\&A, 376, 158

\bibitem[\protect\citeauthoryear{{Allard}, {Hauschildt} \&
  {Schweitzer}}{{Allard} et~al.}{2000}]{allard00}
{Allard} F.,  {Hauschildt} P.~H.,    {Schweitzer} A.,  2000, ApJ, 539, 366

\bibitem[\protect\citeauthoryear{{Byrne}}{{Byrne}}{1992}]{byrne92}
{Byrne} P.~B.,  1992, in NATO ASIC Proc. 375: Sunspots. Theory and Observations
  {Starspots}.
pp 63--73

\bibitem[\protect\citeauthoryear{{Byrne}}{{Byrne}}{1996}]{byrne96}
{Byrne} P.~B.,  1996, in IAU Symposia Vol.~176, On the believability of polar
  spots.
p.~299

\bibitem[\protect\citeauthoryear{Collier~Cameron}{Collier~Cameron}{1997}]{came%
ron97dots}
Collier~Cameron A.,  1997, MNRAS, 287, 556

\bibitem[\protect\citeauthoryear{{Hatzes}}{{Hatzes}}{1998}]{hatzes98}
{Hatzes} A.~P.,  1998, A\&A, 330, 541

\bibitem[\protect\citeauthoryear{Hatzes \& Vogt}{Hatzes \&
  Vogt}{1992}]{hatzes92eieri}
Hatzes A.~P.,  Vogt S.~S.,  1992, MNRAS, 258, 387

\bibitem[\protect\citeauthoryear{{Hauschildt}, {Allard}, {Ferguson}, {Baron} \&
  {Alexander}}{{Hauschildt} et~al.}{1999}]{hauschildt99}
{Hauschildt} P.~H.,  {Allard} F.,  {Ferguson} J.,  {Baron} E.,    {Alexander}
  D.~R.,  1999, ApJ, 525, 871

\bibitem[\protect\citeauthoryear{{Hilditch}, {McLean} \& {Harland}}{{Hilditch}
  et~al.}{1979}]{hilditch79}
{Hilditch} R.~W.,  {McLean} B.~J.,    {Harland} D.~M.,  1979, MNRAS, 187, 797

\bibitem[\protect\citeauthoryear{{Jeffers}, {Collier Cameron}, {Barnes} \&
  {Donati}}{{Jeffers} et~al.}{2004}]{jeffers04c}
{Jeffers} S.~V.,  {Collier Cameron} A.,  {Barnes} J.~R.,    {Donati} J.~F.,
  2004, MNRAS, submitted

\bibitem[\protect\citeauthoryear{{Lehmann}, {Hempelmann} \& {Wolter}}{{Lehmann}
  et~al.}{2002}]{lehmann02}
{Lehmann} H.,  {Hempelmann} A.,    {Wolter} U.,  2002, A\&A, 392, 963

\bibitem[\protect\citeauthoryear{O'Dell, Hendry \& Collier~Cameron}{O'Dell
  et~al.}{1994}]{odell94distance}
O'Dell M.~A.,  Hendry M.~A.,    Collier~Cameron A.,  1994, MNRAS, 268, 181

\bibitem[\protect\citeauthoryear{{Ol\'{a}h}, {Strassmeier} \&
  {Weber}}{{Ol\'{a}h} et~al.}{2002}]{olah02uzlib}
{Ol\'{a}h} K.,  {Strassmeier} K.~G.,    {Weber} M.,  2002, A\&A, 389, 202

\bibitem[\protect\citeauthoryear{O'Neal, Neff \& Saar}{O'Neal
  et~al.}{1998}]{oneal98tio}
O'Neal D.,  Neff J.,    Saar S.,  1998, ApJ, 507, 919

\bibitem[\protect\citeauthoryear{{Patkos} \& {Hempelmann}}{{Patkos} \&
  {Hempelmann}}{1994}]{patkos94}
{Patkos} L.,  {Hempelmann} A.,  1994, A\&A, 292, 119

\bibitem[\protect\citeauthoryear{Petit, Donati \& Collier~Cameron}{Petit
  et~al.}{2002}]{petit02hr1099}
Petit P.,  Donati J.-F.,    Collier~Cameron A.,  2002, MNRAS, 334, 374

\bibitem[\protect\citeauthoryear{{Popper}}{{Popper}}{1996}]{popper96}
{Popper} D.~M.,  1996, ApJS, 106, 133

\bibitem[\protect\citeauthoryear{Rainger, Hilditch \& Edwin}{Rainger
  et~al.}{1991}]{rainger91xyuma}
Rainger P.~P.,  Hilditch R.~W.,    Edwin R.~P.,  1991, MNRAS, 248, 168

\bibitem[\protect\citeauthoryear{{Schrijver} \& {Title}}{{Schrijver} \&
  {Title}}{2001}]{schrijver01polar}
{Schrijver} C.~J.,  {Title} A.~M.,  2001, ApJ, 551, 1099

\bibitem[\protect\citeauthoryear{Spruit \& Weiss}{Spruit \&
  Weiss}{1986}]{spruit86spots}
Spruit H.~C.,  Weiss A.,  1986, A\&A, 166, 167

\bibitem[\protect\citeauthoryear{Strassmeier}{Strassmeier}{1994}]{strassmeier9%
4}
Strassmeier K.~G.,  1994, A\&A, 281, 395

\bibitem[\protect\citeauthoryear{{Strassmeier}}{{Strassmeier}}{2000}]{strassme%
ier00}
{Strassmeier} K.~G.,  2000, A\&A, 357, 608

\bibitem[\protect\citeauthoryear{{Vogt}, {Hatzes}, {Misch} \&
  {K\"{u}rster}}{{Vogt} et~al.}{1999}]{vogt99hr1099}
{Vogt} S.~S.,  {Hatzes} A.~P.,  {Misch} A.~A.,    {K\"{u}rster} M.,  1999,
  ApJS, 121, 547

\bibitem[\protect\citeauthoryear{{Washuettl}, {Strassmeier} \&
  {Collier-Cameron}}{{Washuettl} et~al.}{2001}]{washuettl01}
{Washuettl} A.,  {Strassmeier} K.~G.,    {Collier-Cameron} A.,  2001, in ASP
  Conf. Ser. 223: 11th Cambridge Workshop on Cool Stars, Stellar Systems and
  the Sun {Latest Doppler images of the RS CVn Binary EI Eridani (CD-ROM
  Directory: contribs/washuet)}.
pp 1308

\bibitem[\protect\citeauthoryear{{Weber} \& {Strassmeier}}{{Weber} \&
  {Strassmeier}}{1998}]{weber98ilhyd}
{Weber} M.,  {Strassmeier} K.~G.,  1998, A\&A, 330, 1029

\end{thebibliography}

\clearpage

\clearpage

\begin{table}
\fontsize{10}{12}\selectfont  
\begin{tabular}{l c c c c c l }
%\begin{tabular}{ l l l l l l } 
\hline
\hline
{Visit}& {Obs. Date} & {UT} & {UT} & {Exposure} & {No of} \\
& & {Start} & {End} & {Time(s)} & {Frames}& \\
\hline

1 & {01 November 2001} & 20:55:56 & 01:00:17 & 30 & 165 \\
2 & {03 November 2001} & 14:34:29 & 18:38:21 & 30 & 165 \\
3 & {05 November 2001} & 09:49:17 & 13:52:22 & 30 & 165 \\

\hline
\hline
\end{tabular}
\caption{HST Observations of SV Cam}
\label{hstobs28}
\end{table}

\clearpage

\begin{table}

\label{par}
\fontsize{10}{12}\selectfont  
\hspace{1cm}
\begin{tabular}{l c c }

\hline
\hline

& {Primary Star} & {Secondary Star}\\

\hline

Radius & {1.24$\pm 0.04$\,R$_\odot$} & {0.79$\pm 0.03$\,R$_\odot$} \\ 
Mass & {1.14$\pm 0.03$\,M$_\odot$} & {0.73$\pm 0.02$\,M$_\odot$} \\
log g & {4.31$\pm 0.04$} & {4.5$\pm 0.03$} \\

\hline
\hline
\end{tabular}
\caption{Stellar Parameters of SV Cam}
\end{table}

\end{document}